%% file: conference_101719.tex
\documentclass[conference, a4paper]{IEEEtran}
\IEEEoverridecommandlockouts
\usepackage{cite}
\usepackage{array}
\usepackage{amsmath,amssymb,amsfonts}
\usepackage{algorithmic}
\usepackage{graphicx}
\usepackage{tabularx}
\usepackage{booktabs}
\usepackage{setspace}
\usepackage{multirow}
\usepackage{fancyhdr}
\usepackage{float}
\usepackage{textcomp}
\usepackage{xcolor}
\def\BibTeX{{\rm B\kern-.05em{\sc i\kern-.025em b}\kern-.08em
    T\kern-.1667em\lower.7ex\hbox{E}\kern-.125emX}}
\begin{document}

\title{Coupling Matrix-based Beamforming for Superdirective Antenna Arrays
}

\author{\IEEEauthorblockN{Liangcheng Han\IEEEauthorrefmark{1},
Haifan Yin\IEEEauthorrefmark{1}, Thomas L. Marzetta\IEEEauthorrefmark{2}}
\IEEEauthorblockA{\IEEEauthorrefmark{1}Huazhong University of Science and Technology, Wuhan, China \\\IEEEauthorrefmark{2}Tandon School of Engineering, New York University, Brooklyn, NY\\
Email: \IEEEauthorrefmark{1}hanlc@hust.edu.cn,
\IEEEauthorrefmark{1}yin@hust.edu.cn,\IEEEauthorrefmark{2}tom.marzetta@nyu.edu}\thanks{This work was supported by the National Natural Science Foundation of China under Grant 62071191.}}

\maketitle
\begin{abstract}
In most multiple-input multiple-output (MIMO) communication systems, e.g., Massive MIMO, the antenna spacing is generally no less than half a wavelength. It helps to reduce the mutual coupling and therefore facilitate the system design. The maximum array gain is the number of antennas in this settings. However, when the antenna spacing is made very small, the array gain of a compact array can be proportional to the square of the number of antennas - a value much larger than the traditional array. To achieve this so-called ``superdirectivity" however, the calculation of the excitation coefficients (beamforming vector) is known to be a challenging problem. In this paper, we derive the beamforming vector of superdirective arrays  based on a novel coupling matrix-enabled method. We also propose an approach to obtain the coupling matrix, which is derived by the spherical wave expansion method and active element pattern. The full-wave electromagnetic simulations are conducted to validate the effectiveness of our proposed method. Simulation results show that when the beamforming vector obtained by our method is applied, the directivity of the designed dipole antenna array has a good agreement with the theoretical values. 

\end{abstract}


\section{Introduction}
\addtolength{\topmargin}{0.01in}
As one of the key technologies of 5G, Massive MIMO is being commercialized in the fifth generation (5G) wireless communication systems\cite{marzetta2016fundamentals}. The pioneering paper\cite{marzetta2010noncooperative} of Massive MIMO predicted the large gain of the spectral efficiency when the number of antennas at the base station tends to infinity. However, in practice, the antenna spacing is generally no less than half a wavelength. One of the major reasons is to make the coupling effect between antennas negligible and therefore the system design is simpler. However it also leads to the limitation of the number of antennas for an antenna panel with a fixed size. Moreover, the coupling effect is not all negative. In a compact array containing $M$ antennas with much smaller antenna spacing than half a wavelength, the strong coupling is the foundation of the superdirectivity, where the beamforming gain may reach $M^2$, instead of $M$ as in traditional MIMO theory. 


In recent years, with the further demand for spectral efficiency, the possibility of deploying super dense antenna arrays at the base station has been considered by researchers. 
The brand new concept of Holographic MIMO\cite{bjornson2019massive}\cite{pizzo2021holographic} \cite{9110848} has appeared in the vision of wireless communication researchers. In Holographic MIMO, the locations of antennas is no longer discrete; all antennas will be continuously distributed on the antenna panel. In this concept, the antenna spacing approaches zero, which makes it a strongly coupled antenna array. However, only the channel coupling was studied \cite{pizzo2021holographic}, and few works have considered the antenna coupling so far under this concept. 


The plane wave can be divided into propagating waves and evanescent waves \cite{clemmow2013plane}, and the coupling between antennas is caused by the joint action of propagating waves and evanescent waves. 
When the antenna spacing is much smaller than half a wavelength, the effect of the evanescent wave is more obvious than the propagating wave, making the coupling between antennas extremely strong \cite{hansen2014exact}\cite{altshuler2005monopole}. The role of coupling in antenna array, however, is often ignored in traditional communication research\cite{8861014}. In theory, if the amplitude and phase of each antenna excitation are controlled precisely, this strong coupling can cause the antenna array to have the superdirectivity\cite{bjornson2019massive}\cite{bloch1953new}. 
According to traditional communication theory, the array gain is proportional to the number of antennas $M$. 
However in \cite{uzkov1946approach}, Uzkov has proved that the directivity of an isotropic linear array of $M$ antennas can reach $M^2$ when the spacing between antennas approaches zero. If the base station is equipped with large number of antennas, the array gain improvement is more significant\cite{marzetta2019super}.

In recent years, the realization methods of superdiectivity includes the array theory-based methods\cite{altshuler2005monopole} and the spherical wave expansion methods\cite{clemente2015design}. The maximum directivity of linear dipole array has been analyzed numerically in\cite{altshuler2005monopole} and was shown to be equal to 5.24, 10.8, 18.4 in the case of two, three, four antennas in the array. However, to the best of our knowledge, there is no method to model the coupling effect between antennas in superdiretive array, which cannot be ignored due to the small inter-element spacing. 

In this paper, the closed-form expression of the array directivity is first presented and then the beamforming vector of the superdirective array is derived based on array theory without considering the coupling effects. 
We then propose an approach to obtain the beamforming vector to achieve superdirectivity for a compact array with antenna coupling taken into consideration. The essence of this method is based on the coupling matrix and the spherical wave expansion method. We also propose a method to calculate the coupling matrix with active element pattern. 
Finally, a printed dipole array working in 845 MHz is designed  to validate the effectiveness of our proposed method. The array gains brought by our method matches the theoretical values. 

\section{Superdirective beamforming of antenna arrays}
\addtolength{\topmargin}{0.01in}
Superdirectivity of antenna array is obtained only when at least two conditions are met: first, the spacing between two antennas should be less than half a wavelength, which can be achieved easily. Second, the amplitude and phase of the excitation coefficient of each antenna need to be precisely calculated and controlled.  In this section, the beamforming vector of superdiretive arrays to meet the second condition will be presented. 
\subsection{Beamforming of superdirective arrays based on array theory}

Consider a uniform linear array consisting of $M$ antennas, each with the same pattern function of $k(\theta,\phi)$, where $\theta$ and $\phi$ represent the far-field position in spherical coordinate system. It is assumed the first antenna is located at the origin of the Cartesian coordinate system, while the other antennas are equally spaced by a spacing $d$  on the positive half-axis of the $z$-axis. Hence, the complex far-field pattern $f(\theta,\phi)$ of this antenna array is given by
\begin{align}\label{array_pattern_function}
f(\theta,\phi)=\sum_{m=1}^Ma_mk(\theta,\phi)e^{jk\mathbf{\hat r}\cdot \mathbf{r}_m},
\end{align}
where $a_m$ is the complex excitation coefficient proportional to the current on the $m$-th antenna, $k$ is the wave number, $\mathbf{\hat r}$ is the unit vector of the far-field direction $(\theta,\phi)$ in the spherical coordinate system,  $\mathbf{r}_m$ is the position of the $m$-th antenna. Thus, the  directivity factor at $(\theta_0,\phi_0)$ can be obtained as
\begin{align}\label{theta0_phi0_pattern_function}
  D(\theta_0,\phi_0)=\dfrac{|\sum_{m=1}^{M}a_mk(\theta_0,\phi_0)e^{jk\mathbf{\hat r}_0\cdot\mathbf{ r}_m}|^2}{\frac{1}{4\pi}\int_0^{2\pi}\int_0^\pi |\sum_{m=1}^{M}a_mk(\theta,\phi)e^{jk\mathbf{\hat r}\cdot \mathbf{r}_m}|^2\sin\theta d\theta d\phi},
\end{align}
where $\mathbf{\hat{r}}_0$ is the unit vector in the direction $(\theta_0,\phi_0)$. 
\begin{spacing}{1}
For deriving the beamforming vector of a superdirective array, the expression of \eqref{theta0_phi0_pattern_function} is not intuitive  and is thus worthy of further simplification. The denominator of \eqref{theta0_phi0_pattern_function} can be expanded as
\end{spacing}

\begin{small}
\begin{align}\label{denominator}
&\frac{1}{4 \pi} \int_{0}^{2 \pi} \int_{0}^{\pi}\left|\sum_{m=1}^{M} a_{m} k(\theta, \phi) e^{j k \mathbf{\hat{r}} \cdot \mathbf{r}_m}\right|^{2} \sin \theta d \theta d \phi \notag\\
&=\frac{1}{4 \pi} \int_{0}^{2 \pi} \int_{0}^{\pi} \sum_{n=1}^{N} \sum_{m=1}^{M} a_{n} a_{m}^{*}|k(\theta,\phi)|^2 e^{j k \mathbf{\hat{r}} \cdot \mathbf{r}_n} e^{-j k \mathbf{\hat{r}} \cdot \mathbf{r}_m} \sin \theta d \theta d \phi \notag\\
&= \sum_{n=1}^{N} \sum_{m=1}^{M} a_{n} a_{m}^{*} \frac{1}{4 \pi}\int_{0}^{2 \pi} \int_{0}^{\pi} |k(\theta,\phi)|^2e^{j k \mathbf{\hat{r}} \cdot \mathbf{r}_n} e^{-j k \mathbf{\hat{r}} \cdot \mathbf{r}_m} \sin \theta d \theta d \phi,
\end{align}
\end{small}
where the $a_n^*$ stands for the conjugate of the complex value $a_n$. 

For the integral items in \eqref{denominator}, we introduce the following equation:
\begin{align}\label{zmn}
z_{mn}=\frac{1}{4\pi}\int_0^{2\pi}\int_0^\pi |k(\theta,\phi)|^2e^{jk\mathbf{\hat r}\cdot\mathbf{r}_m }e^{-jk\mathbf{\hat r}\cdot \mathbf{r}_n}\sin\theta d\theta d\phi.
\end{align}
As the power radiated by the antenna array is active power, $z_{mn}$ can represent the real part of the mutual impedance between the $m$-th antenna and the $n$-th antenna.
Then \eqref{denominator} can be rewritten as
\begin{align}\label{simplify_denominator}
  &\sum_{m=1}^{M}\sum_{n=1}^{N}a_ma^*_n\frac{1}{4\pi} \int_0^{2\pi}\int_0^\pi |k(\theta,\phi)|^2e^{jk\mathbf{\hat r}\cdot \mathbf{r}_m }e^{-jk\mathbf{\hat r}\cdot \mathbf{r}_n}\sin\theta d\theta d\phi \notag \\
  &=\sum_{m=1}^{M}\sum_{n=1}^{N}a_ma^*_nz_{mn}.
\end{align}
For the simplicity of notations, \eqref{theta0_phi0_pattern_function} is further rewritten using two vectors $\mathbf{a},\mathbf{e}\in \mathbb{C}^{M\times 1}$
\begin{align}\label{vector_D}
D=\dfrac{\mathbf{a}^T\mathbf{e}\mathbf{e}^H\mathbf{a}^*}{\mathbf{a}^T\mathbf{Z}\mathbf{a}^*},
\end{align}
where $\mathbf{a}^T$  is the transpose of  $\mathbf{a}$ and  $\mathbf{e}^H$ represents the conjugate transpose  of $\mathbf{e}$, $\mathbf{a}$ denotes the beamforming vector
\begin{align}\label{vector_a}
	\mathbf{a}=\left[ a_1,\,a_2,\,\cdots ,\,a_M \right]^T,
\end{align}
and
\begin{align}\label{vector_e}
	\mathbf{e}=\left[ e^{jk\mathbf{\hat{r}}\cdot \mathbf{r}_1}k(\theta,\phi),\,e^{jk\mathbf{\hat{r}}\cdot \mathbf{r}_2}k(\theta,\phi),\,\cdots ,\,e^{jk\mathbf{\hat{r}}\cdot \mathbf{r}_M}k(\theta,\phi) \right]^T.
\end{align}
$\mathbf{Z}\in \mathbb{C}^{M\times M}$ stands for the real part of the normalized impedance matrix\cite{zucker_antenna_1969}
\begin{align}\label{impedance_matrix}
\mathbf{Z}=\left[\begin{array}{ccc}
z_{11} & \ldots & z_{1 M} \\
\vdots & \ddots & \vdots \\
z_{M 1} & \cdots & z_{M M}
\end{array}\right].
\end{align}
\setlength{\abovedisplayskip}{3pt} 
Note that \eqref{vector_D} is in the form of Rayleigh quotient, the directivity optimization problem can be solved by finding the derivative of \eqref{vector_D} with respect to the vector $\mathbf{a}$ as follow:
\begin{equation}\label{derivative}
  \frac{\partial D}{\partial \mathbf{a}}
=\frac{2\mathbf{e}\mathbf{e}^H\mathbf{a}^*}{\mathbf{a}^T\mathbf{Z}\mathbf{a}^*}-2D\frac{\mathbf{Z}\mathbf{a}^*}{\mathbf{a}^T\mathbf{Z}\mathbf{a}^*}.
\end{equation}
Letting the above formula equal to zero yields
\begin{align}\label{forcezero}
  \mathbf{e}\mathbf{e}^H\mathbf{a}^*=D\mathbf{Z}\mathbf{a}^*.
\end{align}
It can be found that \eqref{forcezero} is a generalized eigenvalue problem of the form $\mathbf{A}\mathbf{x}=\lambda \mathbf{B}\mathbf{x}$, where $\mathbf{A}$ and $\mathbf{B}$ are matrices, and $\mathbf{x}$ is the generalized eigenvector of $\mathbf{A}$ and $\mathbf{B}$, while $\lambda$ is the corresponding generalized eigenvalue.  Multiplying both sides of \eqref{forcezero} by $\mathbf{Z}^{-1}$ yields
\begin{align}\label{general}
  \mathbf{Z}^{-1}\mathbf{e}\mathbf{e}^H\mathbf{a}^*=D\mathbf{a}^*.
\end{align}
The above equation has been shown to have only one  solution for the eigenvalue $D$ \cite{zucker_antenna_1969}. Consequently, the only one non-zero eigenvalue of $\mathbf{Z}^{-1}\mathbf{e}\mathbf{e}^H$ is the maximum value of the directivity factor $D_{\max} $. Hence,  \eqref{general} can be rewritten as
\begin{align}\label{generalmax}
  \mathbf{Z}^{-1}\mathbf{e}\mathbf{e}^H\mathbf{a}^*=D_{\max}\mathbf{a}^*.
\end{align}
Since
\begin{align}\label{hh}
  \mathbf{Z}^{-1}\mathbf{e}\mathbf{e}^H\mathbf{a}^*&=\mathbf{Z}^{-1}\mathbf{e}(\mathbf{e}^H\mathbf{a}^*) \notag \\
  &=\xi\mathbf{Z}^{-1}\mathbf{e},
\end{align}
where $\xi=\mathbf{e}^H\mathbf{a}^*$ is a constant, the beamforming vector corresponding to the maximum directivity factor can thus be written as
\begin{align}\label{aa}
  \mathbf{a}=\frac{\xi}{D_{\max}}\mathbf{Z}^{-1}\mathbf{e}^* = \mu \mathbf{Z}^{-1}\mathbf{e}^*,
\end{align}
where the scalar $\mu$ is defined as $\mu = {\xi}/{D_{\max}}$. 
Substituting \eqref{aa} into \eqref{vector_D}, we  obtain the maximum directivity factor as
\begin{align}\label{maxD}
  D_{\max}=\mathbf{e}^H\mathbf{Z}^{-1}\mathbf{e}.
\end{align}

However, the above derivation process ignores  the interaction between the antennas. For example in \eqref{array_pattern_function}, if $a_n$ is set to 1 and $a_m, m=2,\cdots,M, m \ne n$, are set to 0, the $f(\theta,\phi)$ would be $a_n k(\theta,\phi)e^{jk\mathbf{\hat r}\cdot \mathbf{r}_n}$ which indicates that the radiation pattern of the $n$-th antenna is not influenced by any other antennas. Nevertheless, in the superdiretive antenna array, the mutual coupling should not be ignored due to the small antenna spacing. Thus the beamforming vector based on  traditional method \eqref{aa} is not applicable to make the antenna array to produce maximum directivity. Next, we will propose a superdirective array analysing method based on coupling matrix to obtain a more realistic superdirective beamforming vector.

\subsection{Proposed superdirective beamforming based on coupling matrix}
As the superdirectivity of antenna array is obtained only when the antenna spacing is small, the coupling effect  should not be ignored when calculating the excitation. A coupling matrix-based approach is thus proposed in this section to characterize the coupling effect between antennas with the use of the spherical wave expansion method and the full-wave simulation tools.

The spherical wave expansion method is commonly used in antenna measurements to calculate the radiated far-field from the measured spherical near field\cite{hald1988spherical}. It can also decompose an electromagnetic field into a series of spherical wave coefficients. In this method, the electric field $\mathbf{E}(\theta ,\phi )$ radiated by the antenna in the far-field region is expanded as
\begin{align}\label{NE}
\mathbf{E}(\theta ,\phi )&=k\sqrt{\eta}\sum_{s=1}^2{\sum_{n=1}^N{\sum_{m=-n}^n{Q_{s,m,n}}}}\mathbf{K}_{s,m,n}(\theta ,\phi )
\notag\\
&=k\sqrt{\eta}\sum_{s,m,n}{Q_{s,m,n}}\mathbf{K}_{s,m,n}^{}(r,\theta ,\phi ),
\end{align}
where 
\begin{align}
  \mathbf{E}(\theta ,\phi 
)=[E_{\hat{\theta}}(\theta,\phi),E_{\hat{\phi}}(\theta,\phi)],
\end{align}
with $E_{\hat\theta}(\theta,\phi)$ and $E_{\hat\phi}(\theta,\phi)$ being the $\hat\theta$-component and $\hat\phi$-component of the electric field  respectively in the spherical position $\theta$ and $\phi$. $k=2\pi/\lambda$ is the wave number,  and $\eta$ is the medium intrinsic impedance. $N$ is a truncated constant, and the larger $N$ is, the better the above equation \eqref{NE} fits the real electric field. In practice, one may take $N=kr_0+10$, where $r_0$ is the minimum radius of the spherical surface that can enclose the antenna. $Q_{s,m,n}$ is the spherical wave coefficient. $s=1,2$ represents the TE-mode wave and TM-mode wave respectively, $n=1,2,\cdots,N$ is the degree of wave, and $|m|\leq n$ indicates the order of wave. $\mathbf{K}_{s, m ,n}(\theta, \phi)$ is the spherical wave function, which is the solution to the Helmholtz equation with the explicit expression
\begin{align}\label{KK}
\mathbf{K}_{1 ,m ,n}(\theta, \phi)=&[K_{1,m,n}^{(\hat\theta)},K_{1,m,n}^{(\hat\phi)}]\notag\\
=&\sqrt{\frac{2}{n(n+1)}}\left(-\frac{m}{|m|}\right)^{m} e^{j m \phi}(-j)^{n+1} \notag\\
&\left[\frac{j m \bar{P}_{n}^{|m|}(\cos \theta)}{\sin \theta},-\frac{d \bar{P}_{n}^{|m|}(\cos \theta)}{d \theta}\right] \\
\label{KK2}
\mathbf{K}_{2 ,m ,n}(\theta, \phi)=&[K_{2,m,n}^{(\hat\theta)},K_{2,m,n}^{(\hat\phi)}]\notag\\
=& \sqrt{\frac{2}{n(n+1)}}\left(-\frac{m}{|m|}\right)^{m} e^{j m \phi}(-j)^{n} \notag\\
&\left[\frac{d \bar{P}_{n}^{|m|}(\cos \theta)}{d \theta},\frac{j m \bar{P}_{n}^{|m|}(\cos \theta)}{\sin \theta}\right], 
\end{align}
where $\hat{\theta}$ and $\hat{\phi}$ represents the $\hat{\theta}$ component and $\hat{\phi}$ component of spherical unit vector respectively.  $\bar{P}_{n}^{|m|}$ is the normalized associated Legendre function.

The spherical wave expansion can be interpreted as the decomposition of an electric field into a series of orthogonal components with different modes of the  basis for TE and TM waves, respectively.  

In order to calculate the spherical wave expansion coefficients $Q_{s,m,n}$ in \eqref{NE}, the vectorization of \eqref{NE}, \eqref{KK} and \eqref{KK2} is done such that \cite{belmkaddem2015analysis}
\begin{align}\label{barE}
\overline{\boldsymbol{\varepsilon }}= [&E_{\hat\theta}\left( \theta _1,\phi _1 \right) ,E_{\hat\phi}\left( \theta _1,\phi _1 \right) ,E_{\hat\theta}\left( \theta _2,\phi _2 \right) ,E_{\hat\phi}\left( \theta _2,\phi _2 \right), \notag\\&\cdots,
E_{\hat\theta}\left( \theta _P,\phi_P\right) ,E_{\hat\phi}\left( \theta _P,\phi _P \right)  ] ^T
\end{align}
and

\begin{align}\label{barK}
  \overline{{\boldsymbol{K}}}
  =\left(\begin{array}{cccc}
K_{1,-1,1 }^{(\hat\theta)}\left(\theta_{1}, \phi_{1}\right)  & \ldots &K_{2,N,N }^{(\hat\theta)}\left(\theta_{1}, \phi_{1}\right)\\
K_{1,-1,1 }^{(\hat\phi)}\left(\theta_{1}, \phi_{1}\right)  & \ldots &K_{2,N,N }^{(\hat\phi)}\left(\theta_{1}, \phi_{1}\right) \\
K_{1,-1,1 }^{(\hat\theta)}\left(\theta_{2}, \phi_{2}\right)  & \ldots &K_{2,N,N }^{(\hat\theta)}\left(\theta_{2}, \phi_{2}\right) \\
K_{1,-1,1 }^{(\hat\phi)}\left(\theta_{2}, \phi_{2}\right) & \ldots &K_{2,N,N }^{(\hat\phi)}\left(\theta_{2}, \phi_{2}\right)\\
\ldots & \ldots  & \ldots \\
K_{1,-1,1 }^{(\hat\phi)}\left(\theta_{P}, \phi_{P}\right)& \ldots &K_{2,N,N }^{(\hat\phi)}\left(\theta_{P}, \phi_{P}\right)
\end{array}\right),
\end{align}
where $P$ is the number of angular sampling points, $\overline{\boldsymbol{\varepsilon }}$ is thus a column vector of $2P\times 1$ and $\overline{{\boldsymbol{K}}}$ is a matrix of size $2P\times 2N(N+2)$, $E_{\hat\theta}\left(\theta_{n}, \phi_{n}\right)$ and $E_{\hat\phi}\left(\theta_{n}, \phi_{n}\right)$ represent respectively the $\theta$-component and $\phi$-component of the electric field in the direction of $(\theta_n,\phi_n)$, and similarly $K_{s,m,n}^{(\hat\theta)}(\theta_n,\phi_n)$ and $K_{s,m,n}^{(\hat\phi)}(\theta_n,\phi_n)$ also represent different angular components of the spherical wave function. In addition, the series of spherical wave coefficients can be written in the form of a vector as 
\begin{equation}\label{qq}
\overline{\boldsymbol{q}}=\left[ Q_{1,-1,1},Q_{2,-1,1},Q_{1,0,1},Q_{2,0,1},\cdots, Q_{2,N,N}\right] ^T,
\end{equation}
where $\overline{\boldsymbol{q}}$ is a column vector with $2N(N+2)$ elements. Hence, the vectorized representation of \eqref{NE} is given by
\begin{equation}\label{EKQ}
  \overline{\boldsymbol{\varepsilon}}=k\sqrt{\eta} \overline{{\boldsymbol{K}}} \overline{\boldsymbol{q}}.
\end{equation}
Thus, the spherical wave coefficients $\overline{\boldsymbol{q}}$ can be calculated as
\begin{align}\label{qsolution}
  \overline{\boldsymbol{q}}=\frac{1}{k\sqrt{\eta}}\left( \overline{\boldsymbol{K}} \right) ^{\dagger}\overline{\boldsymbol{\varepsilon }}.
\end{align}
The symbol $^{\dagger}$ represents the pseudoinverse operation.

Considering the mutual coupling between antennas,  a new pattern function $l(\theta,\phi)$ of superdirective array based on coupling coefficients can be written as\cite{1330259}
\begin{align}\label{couplefxt}
	l(\theta,\phi)=\sum_{m=1}^{M}\sum_{n=1}^{N}c_{nm}a_mk(\theta,\phi)e^{jk\mathbf{\hat{r}}\cdot\mathbf{r}_n},
\end{align}
where $c_{nm}$ is the coupling coefficient between the $m$-th antenna and the $n$-th antenna. 
For example, if $a_n$ is set to 1 and $a_m, m=1,\cdots,M,m\neq n$ are set to 0, the radiation pattern $l^{(n)}(\theta,\phi)$ of this antenna array will be
\begin{align}\label{ln}
 l^{(n)}(\theta,\phi)=
 k(\theta,\phi)(c_{1n}e^{jk\mathbf{\hat{r}}\cdot\mathbf{r}_1}+\cdots+c_{Mn}e^{jk\mathbf{\hat{r}}\cdot\mathbf{r}_M}).
\end{align}
Thus, the interactions between antennas have been  considered in the coupling coefficients. 

For the convenience of analysis,  the coupling coefficients in the form of a matrix is represented as 
\begin{align}\label{coupling_matrix}
\mathbf{C}=\left[ \begin{matrix}
	c_{11}&		...&		c_{1M}\\
	\vdots&		\ddots&		\vdots\\
	c_{M1}&		...&		c_{MM}\\
\end{matrix} \right].
\end{align}

To calculate the coupling matrix, we propose an approach based on spherical wave expansion and  full-wave simulation. The detailed steps are introduced as follows.

The first step is to obtain the electric field radiated by the antenna array without considering the coupling effect between antennas, i.e., the coupling matrix $\mathbf{C}$ is an identity matrix, and only one antenna of this array is excited. Specifically, a single antenna modeled in full-wave simulation software is first placed on the origin of the coordinate system, and then the electric field $\mathbf{e}_{s1}\in \mathbb{C}^{2P\times 1}$, as shown in \eqref{barE}, is obtained with simulation. Next, the antenna is moved to $[0,0,d]$ where the electric field $\mathbf{e}_{s2}$ is obtained similarly. Following the procedures as above, the antenna moves a distance $d$ each time on the positive half-axis of $z$-axis  to obtain its electric field at different positions until finally the antenna is placed at $[0,0,(M-1)d]$. Consequently, the set of electric field for a single antenna at different positions is given by
 \begin{align}\label{Es}
 	\mathbf{E}_s=\left[\mathbf{e}_{s1},\,\mathbf{e}_{s2},\,,\,\mathbf{e}_{s3},\,\cdots,\,\mathbf{e}_{sM}\right],
 \end{align}
 where $\mathbf{E}_s$ is a matrix of size $2P\times M$. The spherical wave coefficients $\mathbf{q}_{sm}$ of $\mathbf{e}_{sm},m=1,\cdots,M$  can be calculated using \eqref{qsolution}. Define $\mathbf{Q}_s\in \mathbb{C}^{2N(N+2)\times M}$ as
   \begin{align}\label{qs}
 	\mathbf{Q}_s=\left[\mathbf{q}_{s1},\,\mathbf{q}_{s2},\,,\,\mathbf{q}_{s3},\,\cdots,\,\mathbf{q}_{sM}\right].
 \end{align}
 The relationship between \eqref{Es} and \eqref{qs} is
 \begin{align}\label{qssoulution}
 	\mathbf{Q}_s=\frac{1}{k\sqrt{\eta}}\overline{\boldsymbol{K}}^{\dagger}\mathbf{E}_s.
\end{align}

The second step is to obtain the electric field radiated by the antenna array while considering the coupling effect between antennas. We also let only one antenna of this array excited, while the others are terminated with matched loads. Specifically, a uniform linear array consisting of $M$ antennas with the spacing $d$ is modeled in full-wave simulation software, where the $m$-th antenna is placed at $[0,0,(m-1)d]$. Next, only the $m(m=1,\cdots,M)$-th antenna is excited to radiate the electric field $\mathbf{e}_{cm}$. Notice that in full-wave simulation, the mutual coupling between antennas has been considered in $\mathbf{e}_{cm}$, which is the superposition of electric field radiated by all antennas. The spherical wave coefficients of $\mathbf{e}_{cm}$, denoted by $\mathbf{q}_{cm}$, are calculated using \eqref{qsolution}, Define  $\mathbf{Q_{c}}\in \mathbb{C}^{2N(N+2)\times M}$ as
\begin{align}\label{Q}
\mathbf{Q}_c=[\mathbf{q}_{c1},\,\mathbf{q}_{c2},\,\cdots,\,\mathbf{q}_{cM}].
\end{align}
From equation \eqref{qsolution}, it can be seen that a series of spherical wave coefficients corresponds to a radiated electric field. Thus, $\mathbf{Q}_s$ represents the set of electric field radiated by one antenna in coupling-free array, while $\mathbf{Q}_c$ stands for the set of electric field radiated by one antenna in realistic antenna array considering the inter-element coupling effect. Therefore, the coupling matrix $\mathbf{C}$ can be taken for a bridge converting $\mathbf{Q}_s$ to $\mathbf{Q}_c$, and the coupling  coefficients can now be calculated by solving the linear equations
\begin{align}
    \left\{ \begin{array}{c}
	{\mathbf{q}_{\mathbf{c}}}_1=c_{11}{\mathbf{q}_{\mathbf{s}}}_1+c_{21}{\mathbf{q}_{\mathbf{s}}}_2+\cdots +c_{M1}{\mathbf{q}_{\mathbf{s}}}_M\\
	{\mathbf{q}_{\mathbf{c}}}_2=c_{12}{\mathbf{q}_{\mathbf{s}}}_1+c_{22}{\mathbf{q}_{\mathbf{s}}}_2+\cdots +c_{M2}{\mathbf{q}_{\mathbf{s}}}_M\\
	\vdots\\
	{\mathbf{q}_{\mathbf{c}}}_M=c_{1M}{\mathbf{q}_{\mathbf{s}}}_1+c_{2M}{\mathbf{q}_{\mathbf{s}}}_2+\cdots +c_{MM}{\mathbf{q}_{\mathbf{s}}}_M\\
\end{array} \right. .
\end{align}
\addtolength{\topmargin}{0.01in}
Hence, the coupling matrix can be calculated as
\begin{align}\label{Cmatrix}
	\mathbf{C}=\mathbf{Q}_s^{\dagger}\mathbf{Q}_c.
\end{align}

Notice that $\mathbf{C}$ is not a symmetric matrix, i.e., $c_{mn}$ is not necessarily equal to $c_{nm}$. This is because the coupling effect of the $m$-th antenna to the $n$-th antenna and the coupling effect of the $n$-th antenna to the $m$-th antenna is influenced by the number of antennas nearby at their respective positions. 

According to \eqref{vector_D} and \eqref{couplefxt}, the directivity factor based on coupling matrix is thus
\begin{align}\label{coupleD}
	D_{c}=\dfrac{(\mathbf{C}\mathbf{b})^T\mathbf{e}\mathbf{e}^H(\mathbf{C}\mathbf{b})^*}{(\mathbf{C}\mathbf{b})^T\mathbf{Z}(\mathbf{C}\mathbf{b})^*},
\end{align}
where $\mathbf{b}$ is the vector of excitation coefficients that maximizes the directivity $D_C$ based on the coupling matrix method. It is observed that this equation is also in the form of Rayleigh quotient. Thus it is analogous to \eqref{aa} to obtain the beamforming vector  $\mathbf{b}$ as
\begin{align}\label{mutuala}
	\mathbf{b}=\zeta\mathbf{C}^{-1}\mathbf{Z}^{-1}\mathbf{e}^*,
\end{align}
where $\zeta$ is a constant, which is determined by the total power constraint. 

To analyze the effect of ohmic loss on the antenna array, it is assumed that the efficiency of each antenna is $\eta_a$. The normalized loss resistance linked antenna efficiency is\cite{zucker_antenna_1969}
\begin{align}\label{rloss}
    r_\text{loss} = \frac{1-\eta_a}{\eta_a}.
\end{align}
Thus, the gain of the antenna array can be written as
\begin{align}\label{gain}
    G=\frac{(\mathbf{C b})^{T} \mathbf{e e}^{H}(\mathbf{C b})^{*}}{(\mathbf{C b})^{T} (\mathbf{Z}+r_{\text {loss }} \mathbf{I}_{N})(\mathbf{C b})^{*}}
\end{align}
where $\mathbf{I}_{N}$ is an identity matrix of size $N\times N$.

\section{Numerical Results}
\begin{figure}[htbp]
  \centering
  \includegraphics[width=2in]{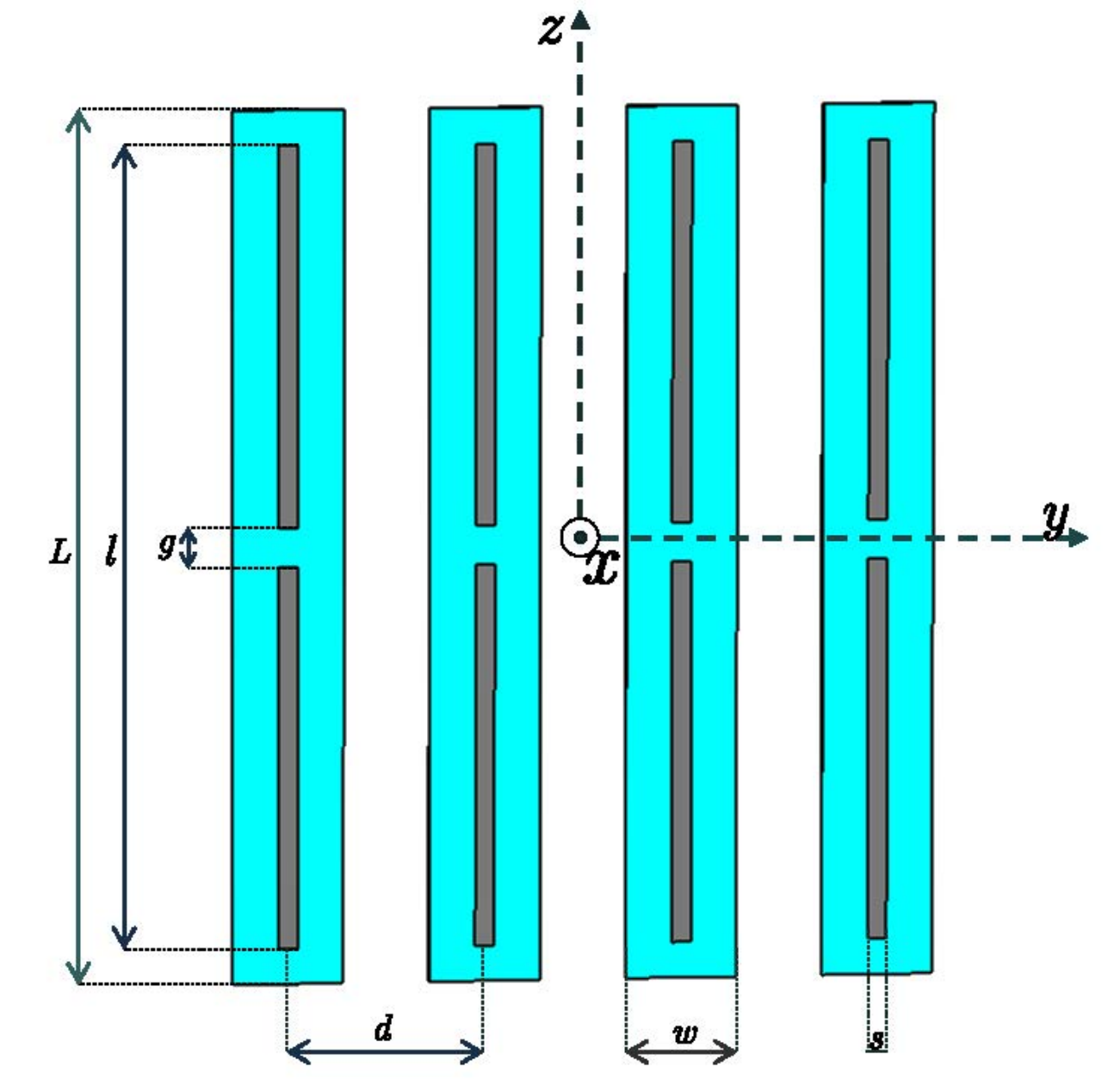}
  \caption{The schematic view of the printed dipole antenna array working in 845 MHz.}\label{fig:1}
\end{figure}

In order to validate the effectiveness of the proposed superdirective beamforming method, full-wave simulations are carried out in this section. A printed dipole antenna array working in 845 MHz is designed as shown in Fig. \ref{fig:1}, where the width and length of the dipole is $w=1\ mm$ and $l=136\ mm$, respectively. The dipole antenna is printed on the FR-4 substrate ($\epsilon_r = 4.47$, $\mu_r=1$ , $\tan\delta= 0.0027$ and the thickness is $813\ \mu m$) of size $30\  mm\times160 \ mm $.  The transmit power is 1 W in simulations.

\begin{figure}[htbp]
  \centering
  \includegraphics[width=2.6in]{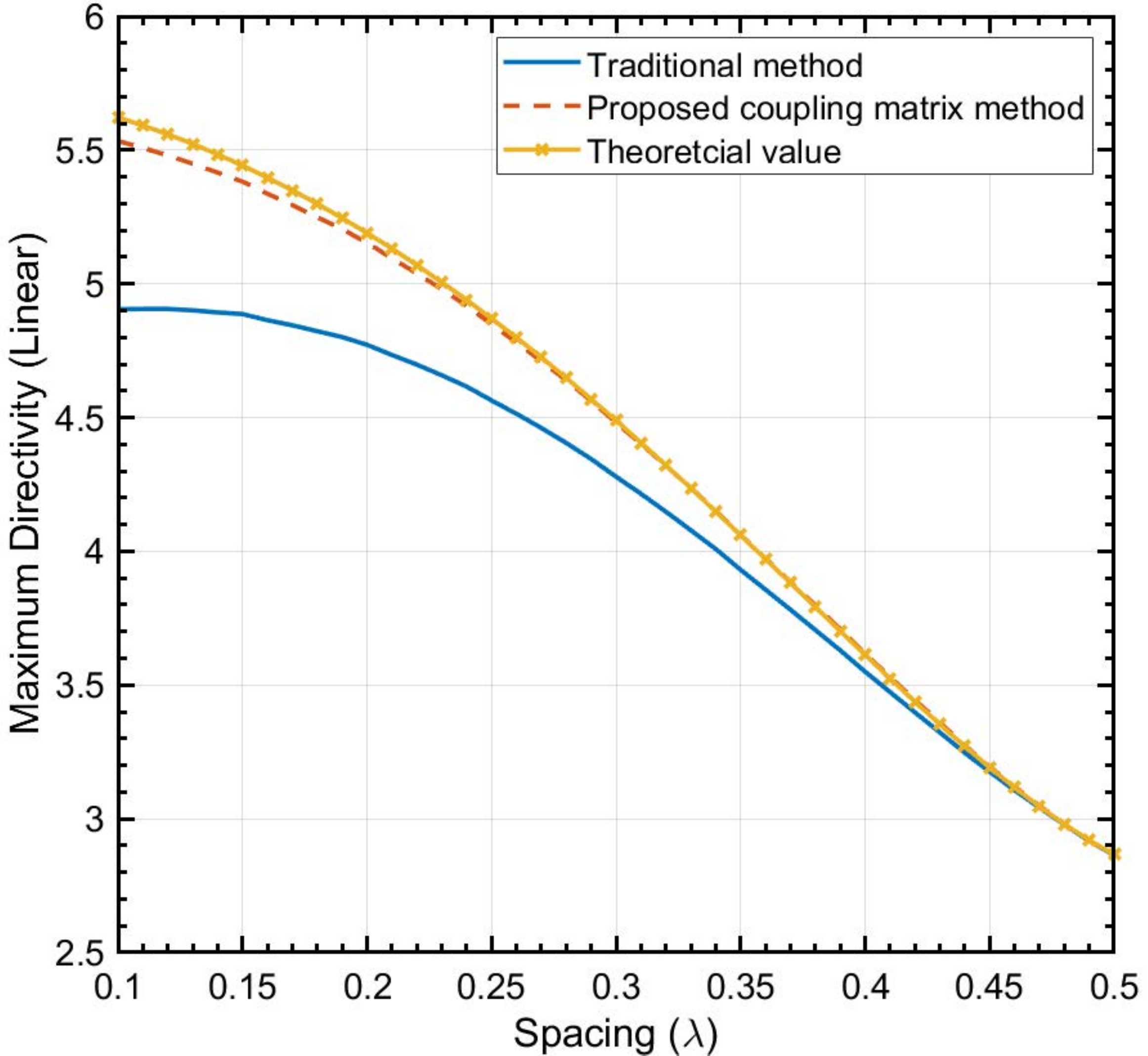}  \caption{The maximum directivity of two dipole antennas excited by the beamforming vector based on traditional method, proposed coupling matrix method, respectively, and the theoretical value. }\label{fig:2}
\end{figure}

\addtolength{\topmargin}{0.01in}
When the number of antennas is two, the simulation results are illustrated in Fig. \ref{fig:2}. The theoretical directivity is calculated by \eqref{coupleD} and \eqref{mutuala}. 
And the directivity curve of the traditional method is obtained when the antenna array is applied with the beamforming vector calculated by \eqref{aa}, which does not consider the mutual coupling between antennas. 
It can be found that the maximum theoretical directivity increases with the decreasing of the antenna spacing, and the directivity reaches $5.6$ when the spacing is $0.1\lambda$. If excited by the beamforming vector based on the traditional method, the directivity of the antenna array also increases as the spacing decreases. However the smaller the spacing, the greater the difference with the theoretical value. 
This is because the coupling effect between antennas is stronger when the spacing is smaller, and the traditional  method does not consider the coupling effect.  However, when excited by the beamforming vector based on our proposed coupling matrix method,  the directivity of antenna array has a good agreement with the theoretical value. When the spacing is around $0.5\lambda$, it can be found that the  three curves almost overlap, which is because the coupling effect can be ignored in this case.

\begin{figure}[htbp]
  \centering
  \includegraphics[width=3.7in]{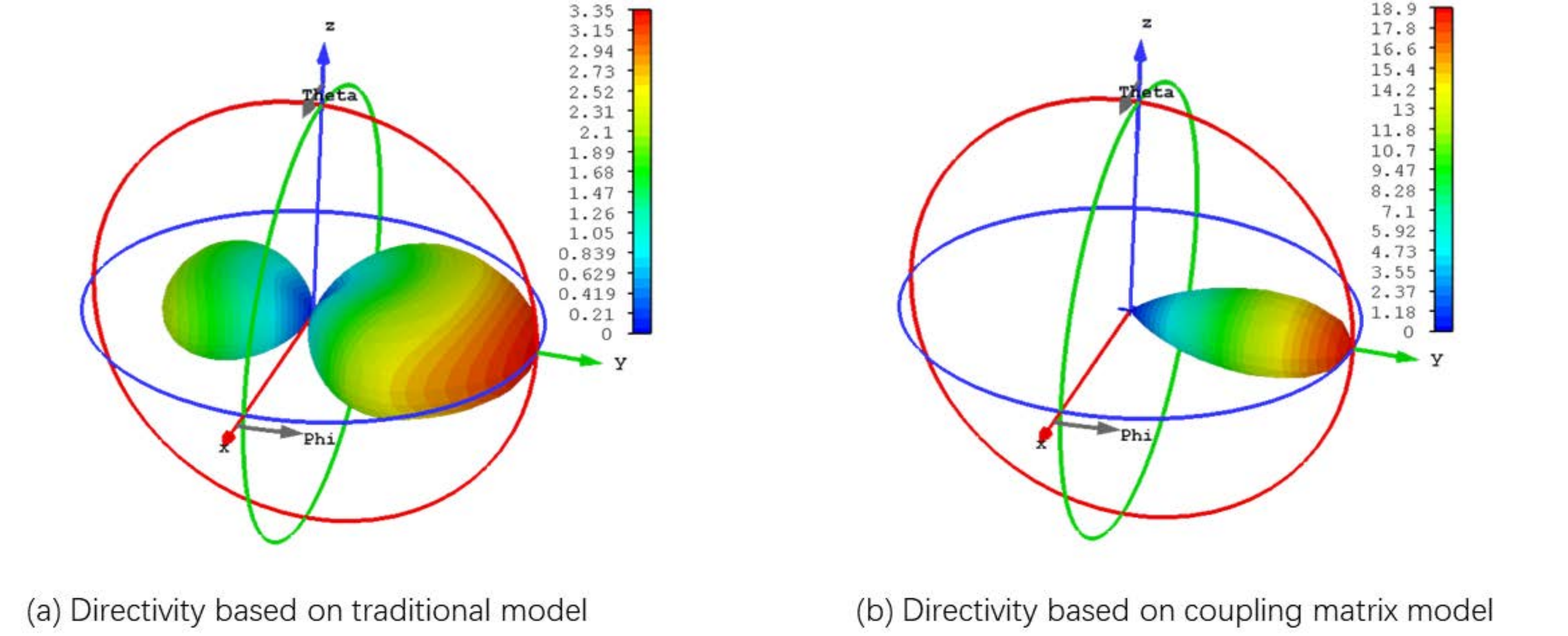}
  \caption{The radiation patterns of the linear dipole antenna arrays, where $M=4$, $d=0.1\lambda$. (a) Excited by optimal excitation coefficients based on traditional method. (b) Excited by the excitation coefficients based on coupling matrix method.  }\label{fig:4}
\end{figure}

Fig. 3 shows the 3D directivity pattern of the linear dipole array with four antennas. The antenna spacing is 0.1$\lambda$. Fig. 3 (a) and Fig. 3 (b) are the patterns when the beamforming vector is calculated by traditional method and by our proposed coupling matrix method respectively. 
It can be seen that the directivity factor of the traditional method is only 3.4, while the theoretical value is 18.9, which is much larger. By introducing the coupling matrix into the model, the 3D directivity pattern  of the antenna array excited by the beamforming vector based on the proposed method is shown in Fig. 3 (b).  
However, when the proposed beamforming method is applied, the directivity is equal to the theoretical value of 18.9. 

\begin{figure}[htbp]
  \centering
  \includegraphics[width=3.2in]{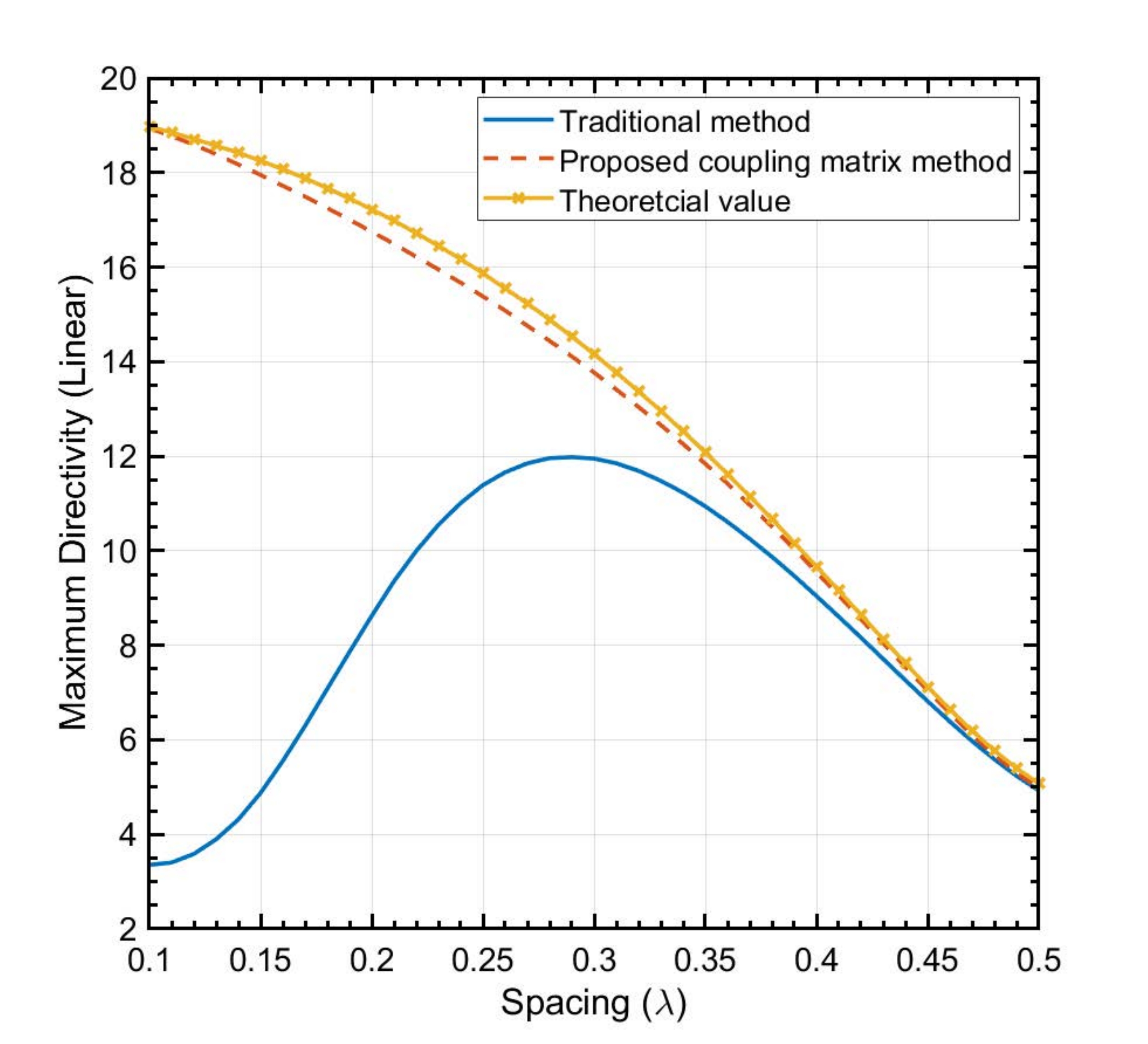}
  \caption{The maximum directivity of four dipole antennas excited by the beamforming vector based on traditional method, proposed coupling matrix method, respectively, and the theoretical value. }\label{fig:3}
\end{figure}

Fig. \ref{fig:3} shows the directivities as a function of the antenna spacing for the four-dipole antenna array. 
It can be seen that the directivity based on traditional method has a poor performance in the small spacing region, while our proposed method is in line with the theoretical values. 

\begin{figure}[htbp]
  \centering
  \includegraphics[width=3in]{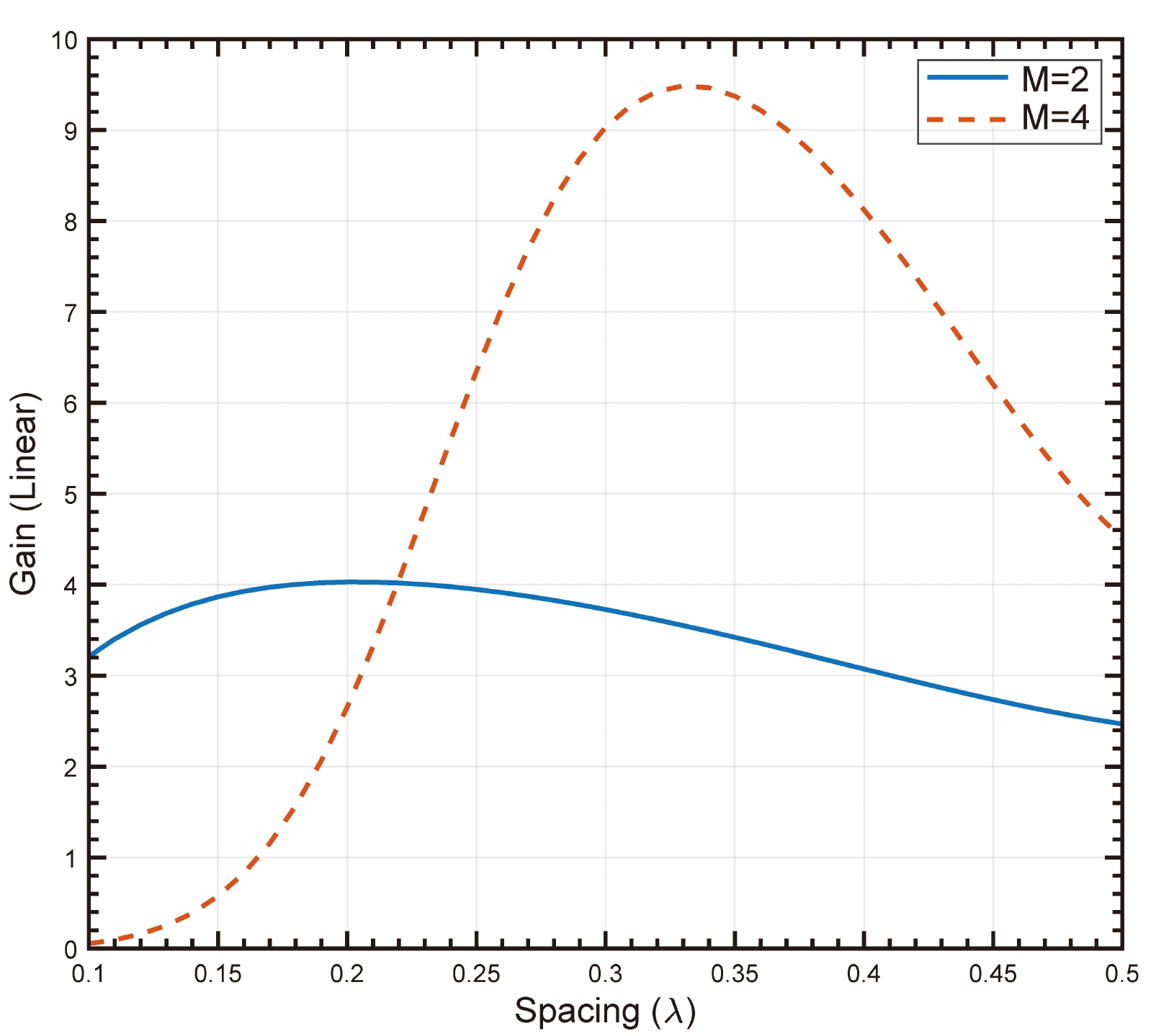}
  \caption{The gain of two and four dipole antennas excited by the beamforming vector based on proposed coupling matrix method, respectively. }\label{fig:5}
\end{figure}


To analyze the effect of ohmic loss on the antenna array, the gain of the antenna array with different spacing is illustrated in Fig. \ref{fig:5}. When the number of antennas is 2 and 4 respectively,  the radiation efficiency of the antenna is 96\%. It can be found the gain does not keep increasing as the spacing decreases, and the larger the number of antennas, the more obvious the effect of ohmic loss at smaller spacing. As a consequence, the maximum gain is 9.6 when the number of antennas is 4 and the spacing is 0.33$\lambda$. This phenomenon can be explained mathematically by \eqref{gain}, where the resistance matrix $\mathbf{Z}$ converges to a singular matrix as the antenna spacing tends to zero. However in this case, the singularity of the resistance matrix changes when a small ohmic loss  $r_{\text{loss}}$ exists, which impairs the antenna array gain significantly. Nevertheless, the problem of ohmic loss can be theoretically alleviated using high-temperature superconducting antennas\cite{hansen2011small}.

\section{Conclusion}
In this paper,  a coupling matrix-based approach to calculate the superdirective beamforming vector is proposed. This method relies on spherical wave expansion method and full-wave simulation method to calculate the coupling matrix precisely. In the simulations, we design a dipole array working in 845 MHz of two and four printed antennas respectively with different antenna spacing to validate the effectiveness of our method. The simulation results show that our method is more accurate than traditional method to make the antenna array to produce maximum directivity. Finally, the ohmic loss impairing the performance of the antenna arrays is analyzed, and the problem can be theoretically alleviated using superconducting antenna.

    \input{conference_101719.bbl}

    \bibliographystyle{IEEEtran}

\end{document}

%% file: conference_101719.bbl

%% file: conference_101719.bbl
\begin{thebibliography}{10}
\providecommand{\url}[1]{#1}
\csname url@samestyle\endcsname
\providecommand{\newblock}{\relax}
\providecommand{\bibinfo}[2]{#2}
\providecommand{\BIBentrySTDinterwordspacing}{\spaceskip=0pt\relax}
\providecommand{\BIBentryALTinterwordstretchfactor}{4}
\providecommand{\BIBentryALTinterwordspacing}{\spaceskip=\fontdimen2\font plus
\BIBentryALTinterwordstretchfactor\fontdimen3\font minus
  \fontdimen4\font\relax}
\providecommand{\BIBforeignlanguage}[2]{{%
\expandafter\ifx\csname l@#1\endcsname\relax
\typeout{** WARNING: IEEEtran.bst: No hyphenation pattern has been}%
\typeout{** loaded for the language `#1'. Using the pattern for}%
\typeout{** the default language instead.}%
\else
\language=\csname l@#1\endcsname
\fi
#2}}
\providecommand{\BIBdecl}{\relax}
\BIBdecl

\bibitem{marzetta2016fundamentals}
T.~L. Marzetta, E.~G. Larsson, H.~Yang, and H.~Q. Ngo, \emph{Fundamentals of
  Massive MIMO}.\hskip 1em plus 0.5em minus 0.4em\relax Cambridge University
  Press, 2016.

\bibitem{marzetta2010noncooperative}
T.~L. Marzetta, ``Noncooperative cellular wireless with unlimited numbers of
  base station antennas,'' \emph{IEEE Trans. Wireless Commun.}, vol.~9, no.~11,
  pp. 3590--3600, 2010.

\bibitem{bjornson2019massive}
E.~Bj{\"o}rnson, L.~Sanguinetti, H.~Wymeersch, J.~Hoydis, and T.~L. Marzetta,
  ``Massive {{MIMO}} is a reality—what is next five promising research
  directions for antenna arrays,'' \emph{Digit. Signal Process.}, vol.~94, pp.
  3--20, 2019.

\bibitem{pizzo2021holographic}
A.~Pizzo, L.~Sanguinetti, and T.~L. Marzetta, ``Holographic {MIMO}
  communications,'' \emph{arXiv e-prints}, pp. arXiv--2105, 2021.

\bibitem{9110848}
A.~Pizzo, T.~L. Marzetta, and L.~Sanguinetti, ``Spatially-stationary model for
  holographic {MIMO} small-scale fading,'' \emph{IEEE J. Sel. Areas Commun.},
  vol.~38, no.~9, pp. 1964--1979, 2020.

\bibitem{clemmow2013plane}
P.~C. Clemmow, \emph{The plane wave spectrum representation of electromagnetic
  fields: International series of monographs in electromagnetic waves}.\hskip
  1em plus 0.5em minus 0.4em\relax Elsevier, 2013.

\bibitem{hansen2014exact}
T.~B. Hansen, ``Exact plane-wave expansion with directional spectrum:
  Application to transmitting and receiving antennas,'' \emph{IEEE Trans.
  Antennas Propag.}, vol.~62, no.~8, pp. 4187--4198, 2014.

\bibitem{altshuler2005monopole}
E.~E. Altshuler, T.~H. O'Donnell, A.~D. Yaghjian, and S.~R. Best, ``A monopole
  superdirective array,'' \emph{IEEE Trans. Antennas Propag.}, vol.~53, no.~8,
  pp. 2653--2661, 2005.

\bibitem{8861014}
L.~Sanguinetti, E.~Björnson, and J.~Hoydis, ``Toward massive {MIMO} 2.0:
  Understanding spatial correlation, interference suppression, and pilot
  contamination,'' \emph{IEEE Trans. Commun.}, vol.~68, no.~1, pp. 232--257,
  2020.

\bibitem{bloch1953new}
A.~Bloch, R.~Medhurst, and S.~Pool, ``A new approach to the design of
  super-directive aerial arrays,'' \emph{Proc. IEE, Pt. III}, vol. 100, no.~67,
  pp. 303--314, 1953.

\bibitem{uzkov1946approach}
A.~Uzkov, ``An approach to the problem of optimum directive antenna design,''
  in \emph{Comptes Rendus (Doklady) de l’Academie des Sciences de l’URSS},
  vol.~53, no.~1, 1946, pp. 35--38.

\bibitem{marzetta2019super}
T.~L. Marzetta, ``Super-directive antenna arrays: Fundamentals and new
  perspectives,'' in \emph{2019 53rd Asilomar Conf. Signals Syst.
  Comput.}\hskip 1em plus 0.5em minus 0.4em\relax IEEE, 2019, pp. 1--4.

\bibitem{clemente2015design}
A.~Clemente, M.~Pigeon, L.~Rudant, and C.~Delaveaud, ``Design of a super
  directive four-element compact antenna array using spherical wave
  expansion,'' \emph{IEEE Trans. Antennas Propag.}, vol.~63, no.~11, pp.
  4715--4722, 2015.

\bibitem{zucker_antenna_1969}
R.~E. Collin, \emph{Antenna theory {Part} 1}, F.~J. Zucker, Ed.\hskip 1em plus
  0.5em minus 0.4em\relax New York: New York, 1969.

\bibitem{hald1988spherical}
J.~Hald and F.~Jensen, \emph{Spherical near-field antenna measurements}.\hskip
  1em plus 0.5em minus 0.4em\relax Iet, 1988, vol.~26.

\bibitem{belmkaddem2015analysis}
K.~Belmkaddem, T.~P. Vuong, and L.~Rudant, ``Analysis of open-slot antenna
  radiation pattern using spherical wave expansion,'' \emph{IET Microw.
  Antennas Propag.}, vol.~9, no.~13, pp. 1407--1411, 2015.

\bibitem{1330259}
S.~Sadat, C.~Ghobadi, and J.~Nourinia, ``Mutual coupling compensation in small
  phased array antennas,'' in \emph{IEEE Antennas Propag. Soc. Int. Symp.,
  2004.}, vol.~4, 2004, pp. 4128--4131 Vol.4.

\bibitem{hansen2011small}
R.~C. Hansen and R.~E. Collin, \emph{Small antenna handbook}.\hskip 1em plus
  0.5em minus 0.4em\relax Wiley Online Library, 2011.

\end{thebibliography}
